\documentclass{aa}
\usepackage{graphics}
\begin{document}
   \title{$uvby-H_{\beta}$ CCD photometry 
          and membership segregation of the open cluster NGC~2548;
	  Gaps in the Main Sequence of open clusters. 
\thanks{Tables~3, 4 and 5 are only available in electronic form at the 
CDS via anonymous ftp to cdsarc.u-strasbg.fr (130.79.128.5) or 
via {\tt http://cdsweb.u-strasbg.fr/cgi-bin/qcat?J/A+A/}}
    }

   \author{L. Ba\-la\-guer-\-N\'u\-\~nez\inst{1,2,3}, 
   C. Jor\-di\inst{1,4}, 
   D. Galad\'{\i}-Enr\'{\i}quez\inst{5}
          }

   \offprints{Balaguer-N\'u\~nez, L., \email{Lola.Balaguer@am.ub.es}}

   \institute{Departament d'Astronomia i Meteorologia, Universitat de
            Barcelona, Avda. Diagonal 647, E-08028 Barcelona, Spain 
         \and
            Shanghai Astronomical Observatory, CAS Shanghai 200030,
                P.R. China
         \and
            Institute of Astronomy, 
	    Madingley Road, CB3 OHA Cambridge, UK
         \and
            CER for Astrophysics, Particle Physics and Cosmology,
	    associated with Instituto de Ciencias del Espacio-CSIC
        \and
            Centro de Astrobiolog\'{\i}a (CSIC-INTA). 
            Carretera de Ajalvir, km 4, E-28850 
            Torrej\'on de Ardoz, Madrid, Spain
             }
\date{Received ; accepted}

\authorrunning{Balaguer-N\'u\~nez et al}
\titlerunning{$uvby-H_{\beta}$ Photometry and Membership of NGC~2548}

   \abstract{
   Deep CCD photometry in the $uvby-H_{\beta}$ intermediate-band system is
   presented for the cluster NGC~2548 (M~48). A complete membership
   analysis
   based on astrometric and photometric criteria is applied.
   The photometric analysis of a selected sample of stars yields a
   reddening value of $E(b-y)$~= 0.06$\pm$0.03, a distance 
   modulus of $V_0-M_V$~= 9.3$\pm$0.5 (725~pc) and a metallicity of
   [Fe/H]~= $-$0.24$\pm$0.27.
   Through isochrone fitting we find an age of $\log t$~= 8.6$\pm$0.1 (400~Myr).
	Our optical photometry and $JHK$ from 2MASS are combined to derive
   effective temperatures of cluster member stars.  
  The effective temperature distribution along 
  the main sequence of the cluster shows several gaps. 
A test to study the significance of these gaps in the main sequence 
of the HR diagram has been applied. 
The method is also applied to several other open clusters (Pleiades, Hyades, 
   NGC 1817 and M 67) to construct a sequence of metallicities and ages. The 
   comparison of the results of each cluster gives four gaps with high 
significance (one of them, centred at 4900~K, has not been previously reported).

      \keywords{
       Galaxy: open clusters and associations: individual: NGC~2548  
       -- Techniques: photometry -- Astrometry -- Methods: observational,
       data analysis, statistical -- Stars: Hertzsprung-Russell (HR) and C-M diagrams}
   }

   \maketitle

%

\section{Introduction}

    The open cluster NGC~2548 (C0811-056), also known as M~48, in Hydra
    ($\alpha_{2000}$=$8^{\mathrm h}13^{\mathrm m}48^{\mathrm s}$, 
    $\delta_{2000}=-5{\degr}48\arcmin$)
    with an estimated distance
    of 630 pc (Pesch \cite{pesch}) or 530 pc (Clari\'a \cite{Claria}), 
    is an intermediate-age open cluster, around $\log t$ = 8.5 (Lyng\aa \ \cite{lynga}),
    with a slightly poorer CN abundance
    than the giants of the Hyades but significantly
    richer than the K giants of the solar neighborhood (Clari\'a
    \cite{Claria}; Twarog et al. \cite{Twa}).
    It has been very poorly studied 
    in spite of being an extended
    object with an apparent diameter of $30\arcmin$ (Trumpler \cite{Trum}) or even 
     $54\arcmin$ (Collinder \cite{Coll}) and brilliant enough to be
    in the Messier list (XVIII century) as number 48 (Messier \cite{mess}).
    It was even considered inexistent for several years 
    due to the fact that Messier quoted its coordinates with an error
    of $5^{\circ}$.
Rider et al.\ (\cite{Rider}) gives photometry in
the Sloan system with a magnitude limit $g'$$\sim$18.
A recent study by Wu et al.\ (\cite{Wu05}) gives BATC photometry in 13 filters.
Radial velocities have been studied by Geyer \& Nelles (\cite{GeNe}) 
with a focal reduced spectrograph. They give data for 23 stars but with a 
very low quality. The only quality measurement is from Wallerstein et al.\
(\cite{Waller}) of star WEBDA~1560.

    This study of NGC~2548 is part of a series of astrometric and photometric
analyses of open clusters that follows already published results on NGC~1817
(Balaguer-N\'u\~nez et al.\ \cite{Bal04a}, \cite{Bal04b}) and that will be
completed with a study of M~67 (Balaguer-N\'u\~nez et al.\ \cite{Bal05}).

    Absolute proper motions of 501 stars within a 1\fdg6 $\times$
    1\fdg6 area in the NGC~2548 region, from automatic MAMA measurements
    of 10 plates taken with the double astrograph at Z\^ o-S\`e station of 
    Shanghai Observatory, were studied by our group 
    (Wu et al. \cite{Wu}, hereafter Paper~I). 


    In this paper we discuss the results of a deep CCD photometric study 
of NGC~2548, covering an area of 
34$^{\prime}$$\times$34$^{\prime}$ down to $V$~$\approx$~22.  
Section~2 contains the details of the CCD observations and their 
reduction and transformation to the standard system. 
In Sect.~3 we discuss a new astrometric 
membership segregation based on the comparison 
between parametric and non-parametric approaches
applied to the proper motions from Paper~I. 
In Sect.~4 we discuss the colour-magnitude diagram
and identify the sample of candidate cluster members
using astrometric as well as photometric criteria. 
Section~5 contains the derivation of the fundamental cluster
parameters of reddening, distance, metallicity and age. 
In Sect.~6 we calculate effective temperatures and study the
significance of gaps in the main sequence. The method is also applied 
to our results on NGC~1817 and NGC~2682 (M~67) 
and to the well-studied Hyades and Pleiades clusters,
allowing us to test the presence of the gaps as a function
of age and metallicity.  
Section~7 summarizes our
conclusions. 


\section{The data}

\subsection{Observations}

        Deep Str\"omgren CCD photometry of the area was performed at the Calar 
Alto Observatory (Almer\'{\i}a, Spain) in January 1999 and January 2000 
using the 1.23 m telescope of the Centro Astron\'omico Hispano-Alem\'an (CAHA)  
and in January 1999 and February 2000 using the 1.52 m telescope of 
the Observatorio Astron\'omico Nacional (OAN).
Further data were obtained at the Observatorio del Roque de los Muchachos
(ORM, La Palma, Canary Islands, Spain) in February 2000 using the 2.5 m 
Isaac Newton Telescope (INT) of ING (equipped with the Wide-Field Camera, WFC), 
and in December 1998 and February 2000 using the 1 m Jakobus Kapteyn 
Telescope (JKT) of ING, with the $H_{\beta}$ filter.
	The poor quality of the images obtained in the 1998/99 runs and 
in the OAN 2000 observations, due to adverse meteorological conditions, 
prevented us from making use of the data collected during those nights.
        A log of the observations with the total number of frames,
exposure times and seeing conditions is given in Table~\ref{log}.

\begin{figure}
\resizebox{\hsize}{!}{\includegraphics{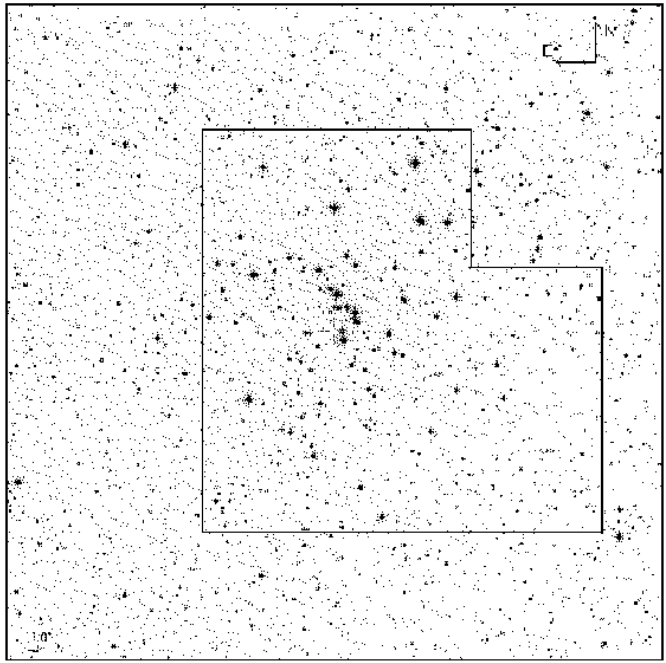}}
\caption{Finding chart of the area under study. The covered area 
         is marked in black on an 
image of a plate plotted with Aladin.}
\label{map}
\end{figure}

 We obtained photometry for a total of 4806 stars in an area of 
34$^{\prime}$$\times$34$^{\prime}$ around NGC~2548, down to a 
limiting magnitude $V$$\sim$22.
The area covered by the observations is shown in the finding chart of 
the cluster (Fig.~\ref{map}). 
	Due to the lack of $H_{\beta}$ filter at the WFC-INT, it was only 
possible to measure it at the JKT and CAHA telescopes, thus 
limiting the spatial coverage with this filter. Only 283 stars have 
$H_{\beta}$ values. Of those only 253 have values of the rest of the
Str\"omgren filters.

\begin{table*}
\begin{center}
\caption {Log of the observations}
\begin {tabular} {ccccccccc}
\hline
   Telescope  & Date & Seeing($\arcsec$) & n. of frames & & Exp. & Times & ($\sec$) & \\
   &  &  &  &  $u$ & $v$ & $b$ & $y$ & $H_\beta$  \\
\hline
  1.23~m CAHA  & 1999/01/12-15 & (1) & 13 & 1900 &  800 & 400 & 400 & 2000 \\

  1.23~m CAHA  & 2000/01/05-10 & 1.1 & 21 & 2200 & 1400 & 900 & 800 & 1400 \\
  1.52~m OAN   & 1999/01/13-16 & (1) & 15 & 1900 & 800 & 400 & 400 & 2000 \\
  1.52~m OAN   & 2000/02/07-14 & (1) & 5 & - & - & - & - & 1400 \\
  1~m JKT &  1998/12/11-14 & (1) & 26 & 2000 & 1200 & 800 & 700 & 1200 \\
  1~m JKT &  2000/02/02-06 & 1.1 & 18 & - & - & - & - & 2000 \\
  2.5~m WFC-INT & 2000/02/02-03 & 1.3 & 17 & 2000 & 2000 & 1200 & 500 & - \\
\hline
\label{log}
\end {tabular}
\end{center}
(1) Poor weather conditions. Images not used in the final data.
\end {table*}

	Beside long, deep exposures, additional shorter exposures
were obtained in order to avoid saturation of the brightest stars.

\subsection{Reduction and transformation to standard system}

The reduction of the photometry is explained at length in 
Balaguer-N\'u\~nez et al. (\cite{Bal04b}). 
Our general procedure has been to routinely obtain twilight sky flats for all 
the filters 
and a sizeable sample of bias frames (around 10) before and/or after every run. 
Flat fields are typically fewer in number, from five to ten per filter.  
Two or three dark frames of 2000 s were also taken. 
IRAF\footnote{IRAF is distributed by the National Optical Astronomy 
Observatories,
which are operated by the Association of Universities for Research
in Astronomy, Inc., under cooperative agreement with the National
Science Foundation.} routines were used for the reduction process.
The bias level was evaluated individually for each frame by averaging the counts
of the most stable pixels in the overscan areas. The 2-D structure of the bias
current was evaluated from the average of a number of dark frames with zero exposure time.
Dark current was found to be negligible in all the cases.
Flatfielding was performed using 
sigma clipped, median stacked, dithered twilight flats.

Our fields are not crowded. Thus, synthetic aperture
techniques provide the most efficient measurements of relative fluxes within 
the frames and from frame to frame. We used the appropiate IRAF packages, 
and DAOPHOT and DAOGROW algorithms (Stetson \cite{Stet87}, \cite{Stet90}). 
We analyzed the magnitude growth curves and determined the aperture correction 
with the IRAF routine MKAPFILE. 

For the WFC images from the INT, we employed  
the pipeline specifically 
developed by the Cambridge Astronomical Survey Unit. 
The process bias subtracts, gain corrects and flatfields the images.
Catalogues
are generated using algorithms described in Irwin (\cite{Irw}).
The pipeline gives accurate positions 
in right ascension and declination linked to the USNO-2 Catalogue 
(Monet et al.\ \cite{USNO2}), and instrumental magnitudes with 
their corresponding
errors. A complete description can be found in Irwin \& Lewis (\cite{Irwin})
and in {\tt http://www.ast.cam.ac.uk/\~{}wfcsur/index.php}.

        The coefficients of the instrumental-to standard 
transformation equations were computed 
by a least squares method using the instrumental magnitudes of the
standard stars and their standard magnitudes and colours in the 
$uvby-H_{\beta}$ system. Up to 68 standard stars from the cluster 
M~67 (Nissen et al.\ \cite{Nissen}) were observed depending on 
the size of the field. Four to six short exposures in every 
filter were taken every night with a magnitude limit of $V$=18. 
Those standard stars with residuals greater than 
2$\sigma$ were rejected. Typically that involves a 10-15\% of the
standard stars, mainly variables from the M~67 field. 
The reduction was performed for each night independently and in two steps.
The first step is to determine the extinction coefficients for
each passband from the standard stars. With the extinction coefficients 
fixed, the transformation coefficients to the standard system 
were fitted.

The mean errors as a function of apparent visual magnitude
are given in Table~\ref{error} for the NGC~2548 stars.

\begin{table*}
\begin{center}
\caption {Number of stars observed ($N$) and mean internal errors 
           ($\sigma$) as a function of apparent visual magnitude.}
\begin {tabular} {ccccccccccc}
\hline
\hline
\multicolumn{1}{c}{$V$ range} & 
\multicolumn{2}{c}{$V$} & 
\multicolumn{2}{c}{$(b-y)$} & 
\multicolumn{2}{c}{$m_1$} & 
\multicolumn{2}{c}{$c_1$} & 
\multicolumn{2}{c}{$H_{\beta}$} \\
\hline
 & $N$&$\sigma$ & $N$&$\sigma$ & $N$&$\sigma$ & $N$&$\sigma$ &$N$&$\sigma$ \\
\hline
 8- 9 &    3 & 0.019 &    3 & 0.028 &    2 & 0.016 &    2 & 0.002 &    2 & 0.017  \\
 9-10 &   19 & 0.012 &   19 & 0.011 &   19 & 0.013 &   16 & 0.019 &    5 & 0.003  \\
10-11 &   25 & 0.014 &   25 & 0.016 &   25 & 0.016 &   24 & 0.010 &    6 & 0.009  \\
11-12 &   32 & 0.012 &   31 & 0.013 &   31 & 0.018 &   31 & 0.016 &   10 & 0.030  \\
12-13 &   45 & 0.012 &   43 & 0.013 &   43 & 0.016 &   43 & 0.011 &   12 & 0.017  \\
13-14 &   71 & 0.012 &   71 & 0.015 &   69 & 0.020 &   69 & 0.017 &   17 & 0.027  \\
14-15 &  152 & 0.012 &  152 & 0.023 &  146 & 0.029 &  146 & 0.030 &   29 & 0.016  \\
15-16 &  260 & 0.015 &  260 & 0.034 &  259 & 0.047 &  251 & 0.047 &   42 & 0.025  \\
16-17 &  380 & 0.022 &  380 & 0.043 &  376 & 0.061 &  360 & 0.063 &   53 & 0.033  \\
17-18 &  514 & 0.030 &  514 & 0.035 &  501 & 0.043 &  468 & 0.043 &   51 & 0.036  \\
18-19 &  639 & 0.027 &  639 & 0.028 &  592 & 0.031 &  477 & 0.032 &    8 & 0.046  \\
19-20 &  752 & 0.012 &  752 & 0.015 &  702 & 0.022 &  383 & 0.038 &      &        \\
20-21 &  859 & 0.023 &  859 & 0.028 &  662 & 0.048 &  160 & 0.052 &      &        \\
21-22 &  763 & 0.049 &  763 & 0.063 &  262 & 0.089 &   21 & 0.082 &      &        \\
22-23 &  222 & 0.095 &  222 & 0.122 &   28 & 0.139 &      &       &      &        \\
\hline
Total & 4736 &       & 4733 &       & 3717 &       & 2451 &       &  235 &        \\
\hline
\end {tabular}
\label{error}
\end{center}
\end {table*}

Table~3 lists the $u,v,b,y,H_{\beta}$ data for all 4806 stars in a
region of 34$^{\prime}$$\times$34$^{\prime}$ around the
open cluster NGC~2548 (Fig.~\ref{map}). Star centres are given in the frame ($x,y$)
and equatorial ($\alpha_{\mathrm J2000}$,$\delta_{\mathrm J2000}$) coordinates. 
An identification number was assigned to each star following 
the order of increasing right ascension.
Column 1 is the ordinal star number; columns 2 and 3
are $\alpha_{\mathrm J2000}$ and $\delta_ {\mathrm J2000}$;
columns 4 and 5 are the respective $x$, $y$ coordinates in arcmin;
columns 6 and 7 are the $(b-y)$ and its error,
8 and 9 the $V$ magnitude and its error, 10 and 11 the $m_1$ and its error, 
12 and 13 the $c_1$ and its error, 
and 14 and 15 the $H_{\beta}$ and its error.
In column 16, stars considered candidate members (see Sect.~4.1.) are labelled 
 'M', while those classified as non-members show the label 'NM'.

\addtocounter{table}{1}

    The cross-identification of stars in common 
    with the astrometry (Paper~I), WEBDA
({\tt http://obswww.unige.ch/WEBDA}), 
Hipparcos (ESA, \cite{esa}), Tycho-2 (H\a{o}g et al.\ \cite{tyc2a}) 
and USNO-2 (Monet et al.\ \cite{USNO2}) catalogues
    is provided in Table~4. 
\addtocounter{table}{1}

\subsection{Comparison with previous photometry}

	Only three stars in the NGC~2548 area have been previously
studied using Str\"omgren photometry (WEBDA 0366, 1560, 2156). 
Unfortunately, none of them is inside the area covered by our 
photometry. Only one
of them (WEBDA 1560) is considered a cluster member from our astrometry 
and will be studied in Sect.~5.
 
The $V$ magnitude derived from the $y$ filter
can be compared with the published broadband data. There are 30 stars
in common with
Pesch (\cite{pesch}). 
The corresponding 
mean difference in $V$, in the sense ours minus Pesch' is $-$0.01$\pm$0.03.

Transformations between
$B-V$ and $b-y$ from several authors (see Moro \& Munari, \cite{ADPS}) fails 
to cover the whole range under study. 
We can find a linear relation between the two indices: 
$B-V$~=~(1.632$\pm$0.031)$(b-y)-(0.038\pm0.010)$, $N$=30.
The standard deviation of 
the residuals about the mean relation is 0.048, where the typical 
uncertainty in $B-V$ is 0.02 and in $b-y$ is 0.016.


\section{Astrometric analysis}

	Paper~I gives absolute proper motions of 501 stars within 
a 1\fdg6 $\times$ 1\fdg6 area in the NGC~2548 region up to  
$B_{\rm inst}$$\approx$14, 
from automatic MAMA measurements of 10 plates and a maximum epoch 
difference of 82 years. Membership determination is calculated in Paper~I
with a 9-parametric Gaussian model and a list of stars with a probability 
higher than 0.7 gives 165 cluster members.  

	NGC~2548 is a very extended object with a complex structure 
with a double core, prolate shape and a tidal tail with a clump
(Bergond et al. \cite{Bergond}). It has been suggested that 
this secondary clump is associated with the last strong disk shock that 
occurred between 20 and 40 Myr ago. Confirmation of members at large radii
will trace the distribution of stars that are currently leaving the
cluster. This will help to constrain models of the tidal disruption of open
clusters. 

 	To complement the cluster/field segregation analysis of the astrometry 
from Paper~I we have applied a non-parametric method to the proper motion data, as
explained in Balaguer-N\'u\~nez et al. (\cite{Bal04a}). In the non-parametric
approach we are able to differenciate the cluster population without the 
need for any a priori knowledge. Following Galad\'{\i}-Enr\'{\i}quez et 
al.\ (\cite{gala2}),
we can perform an empirical
determination of the probability density functions (PDFs) evaluating the 
observed local density in each node of a two-dimensional grid in the vector 
point diagram (VPD), with
a normal circular kernel. The smoothing parameter $h$ 
(Gaussian dispersion) is chosen using Silverman's rule (\cite{Sil}). 
The procedure was tested for several subsamples applying different 
proper motion cutoffs. Satisfactory results are obtained with a proper 
motion cutoff of $|\mu| \leq$ 15 mas~yr$^{-1}$.

The empirical frequency function determined from the 
VPD corresponding to the area occupied by the cluster is made up of two
contributions: cluster and field. To differenciate the two populations we
need to estimate the field distribution.
For this purpose, we studied the VPD for the plate area outside a circle
centered on the cluster. The centre of the cluster was chosen as the point
of highest spatial density. We did tests with circles of very different
radii, searching a reasonable tradeoff between cleanness (absence 
of a significant number of cluster members) and signal-to-noise ratio
(working area not too small). The kernel density estimator was 
applied in the VPD to these data, yielding the empirical frequency function,
for a grid with cell size of 0.2 mas~yr$^{-1}$, well below the proper 
motion errors.

\begin{figure}
\begin{center}
\resizebox{6.5cm}{!}{\includegraphics{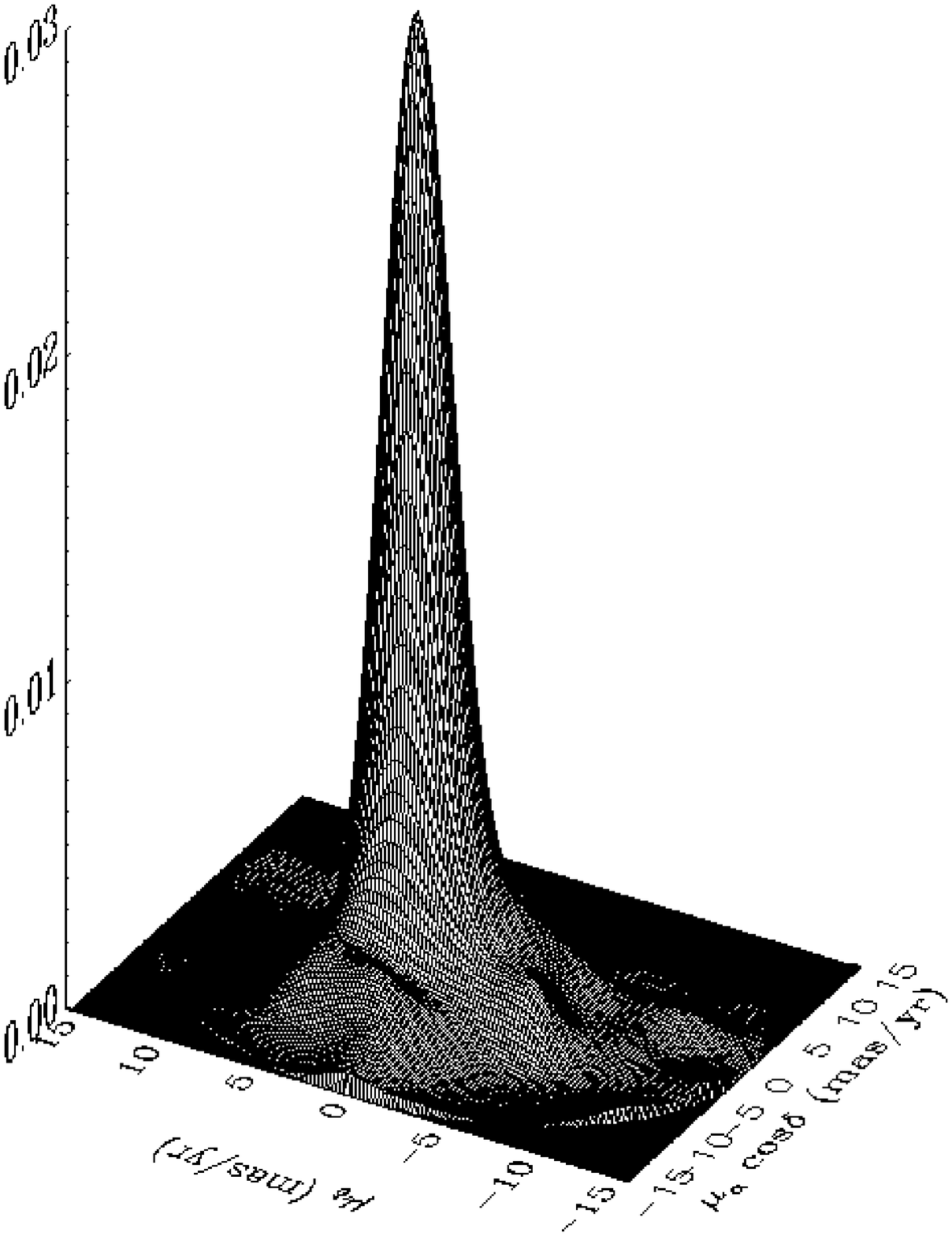}}\\
\resizebox{6.5cm}{!}{\includegraphics{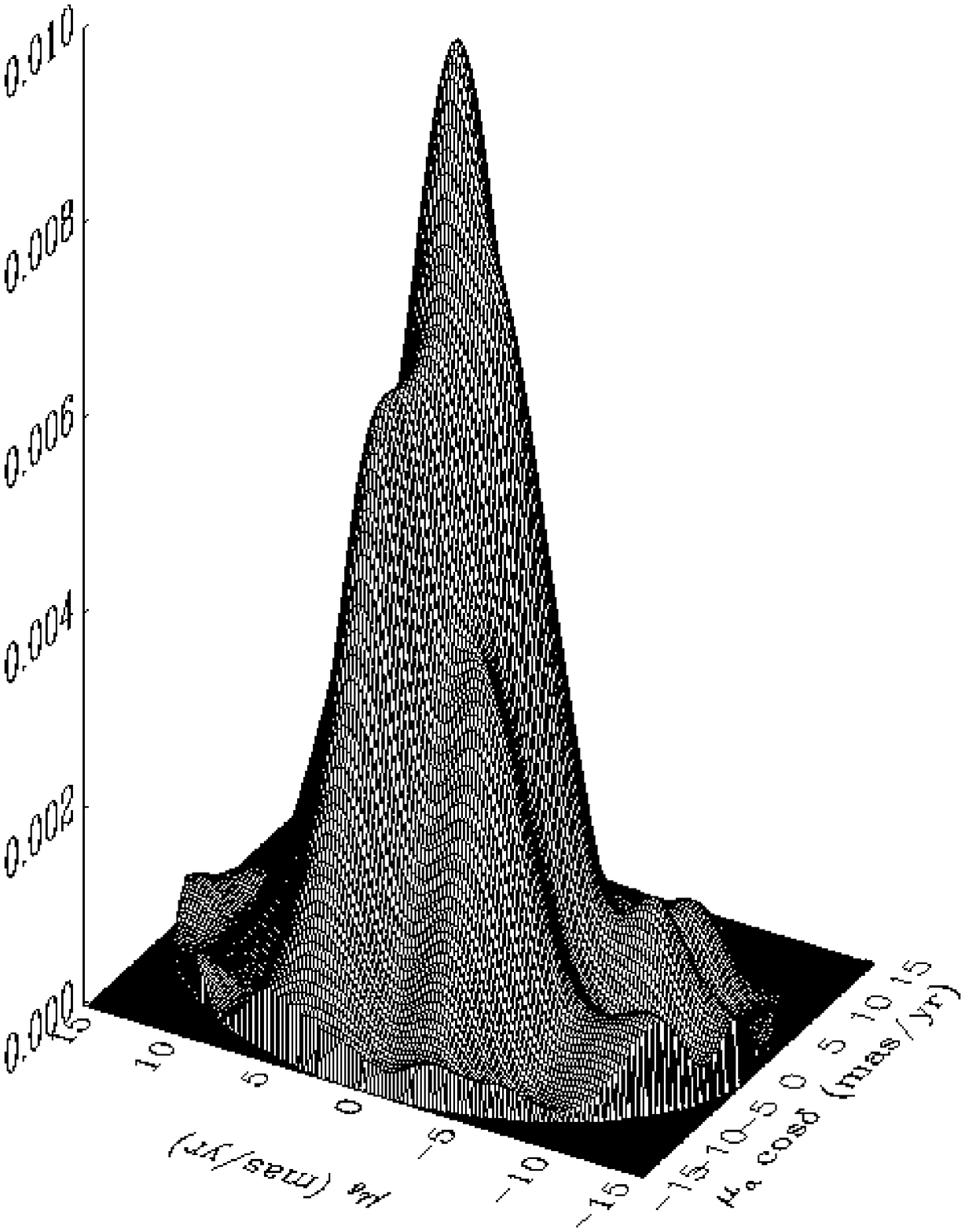}}\\
\resizebox{6,5cm}{!}{\includegraphics{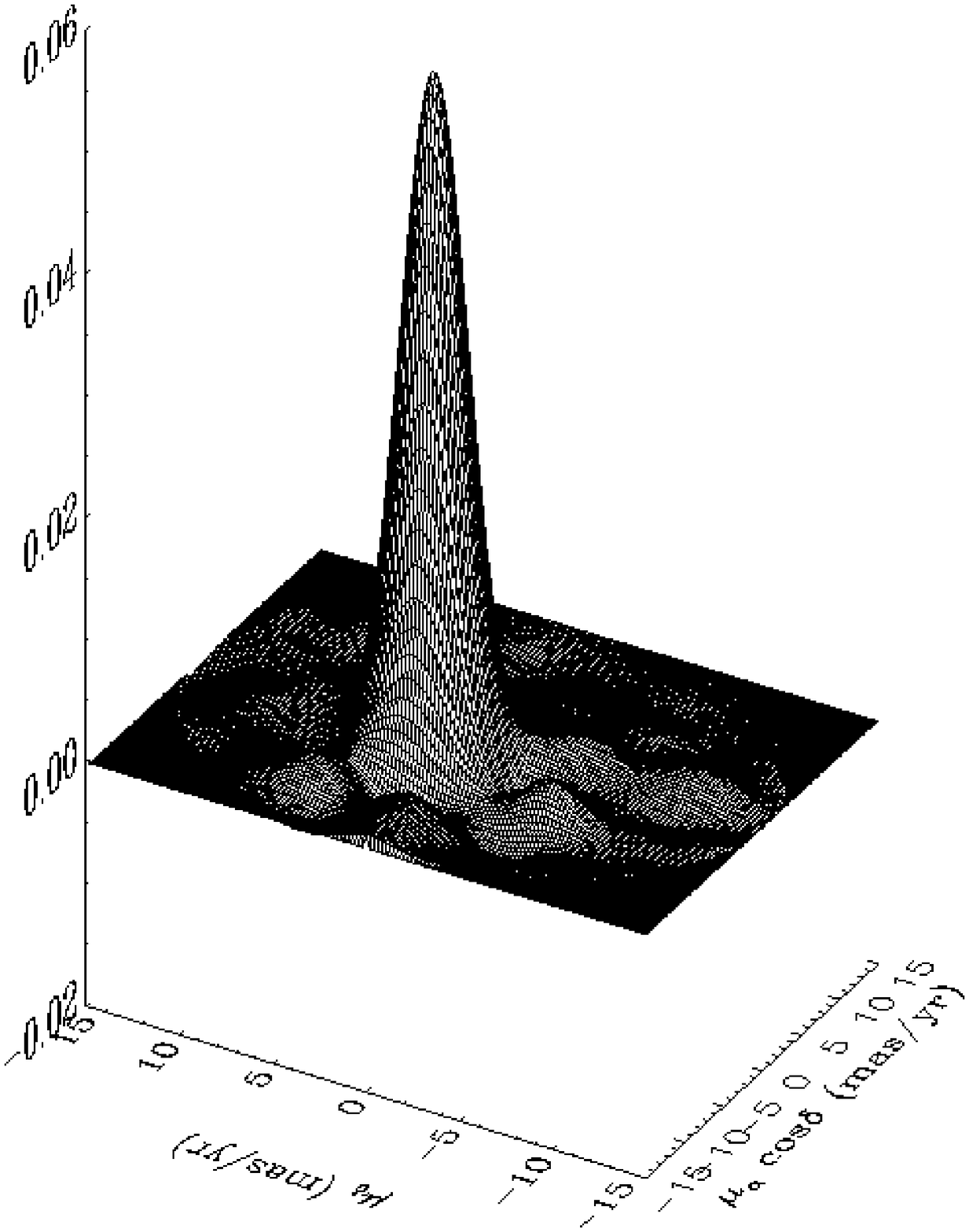}}
\caption{Empirical probability density functions in the kinematic
plane. Top: mixed sample
from the inner circle of 35'. Center: field population from 
outside this circle. Bottom: cluster population of NGC~2548}
\label{nprob}
\end{center}
\end{figure}

We finally found that the area outside a circle with a radius of 
35$\arcmin$ centered on the cluster yields a clean frequency 
function with low cluster contamination and low noise.
The empirical frequency functions can be normalized to yield the 
empirical PDFs for the mixed population (circle), for the field 
(outside the circle) and for the cluster (non-field) population. 
Figure~\ref{nprob} displays these three functions.
Of course, the field PDF estimated in the outer area cannot be an absolutely
perfect representation of the true field PDF in the whole area. This 
introduces undesired noise in the frequency function of the cluster, 
as shown by 
the negative density values found in several zones and by some positive 
fluctuations without physical meaning. These negatives values 
allow us to estimate the typical noise level, $\gamma$, present in the result.
To avoid meaningless probabilities in zones of low density, we restricted the
probability calculations to the stars with cluster PDF $\geq$ 3$\gamma$.
The maximum of the cluster PDF is located at 
($\mu_{\alpha}\cos\delta, \mu_{\delta}$)~= ($-$1.2$\pm$0.2,2.2$\pm$0.2) mas~yr$^{-1}$.

	However, this cluster proper motion is different to the
value found in Paper~I. As discussed by many authors (Galad\'{\i}-Enr\'{\i}quez
et al.\ \cite{gala2} for example) one of the limitations
of the parametric approach is the trend of the circular Gaussian distribution,
used to fit the cluster, to assume an excessive width to improve the
representation of the field distribution. The cluster mean proper motion 
will then be thus affected. To measure the influence of this effect,
we decided to apply the parametric method but to fix the internal velocity
dispersion of the cluster at zero. This way the model will assign to the 
cluster Gaussian distribution a width related only to measurement errors.
We then obtained a mean proper motion of  
($\mu_{\alpha}\cos\delta, \mu_{\delta}$)~= ($-$1.10$\pm$0.08,2.09$\pm$0.08) mas~yr$^{-1}$,
in agreement with the cluster proper motion
obtained by the non-parametric approach. 
 
The non-parametric technique does not take into account the
errors of the individual proper motions, 
therefore it does not make any particular distinction between bright or faint 
stars, different epoch spread and so on. 
However, the FWHM of the empirical cluster PDF provides an estimation of the 
errors of the distribution. We obtained a FWHM of $\sim$4.2$\pm$0.2 
mas~yr$^{-1}$. If the Gaussian dispersion owing to the smoothing 
parameter $h =$~1.44 mas~yr$^{-1}$ is taken into account, 
this FWHM corresponds to a value of 1.53~mas~yr$^{-1}$. 
But from Paper~I we know that the mean proper motion precision is 1.18~mas~yr$^{-1}$
which gives us an intrinsic dispersion component of 0.97~mas~yr$^{-1}$,
(3~km~s$^{-1}$ at the distance of 725~pc from Sect.~5), of the same 
order but slightly lower than the value obtained by the 
membership determination in Paper~I. 
This indicates that the intrinsic velocity dispersion of the cluster cannot be
neglected. 
Although fixing it to zero improves the determination of the mean
motion of the cluster (position of the centre of the fitted Gaussian), 
the membership probability results are more meaningful
taking into account the intrinsic dispersion. In our analysis, we will use the
parametric results from Paper~I. However, slight differences in the center of
the adopted Gaussian do not affect the segregation, since the 
stars with highest probability of being members are almost
the same in both cases.

\begin{figure}
\begin{center}
\resizebox{\hsize}{!}{\includegraphics{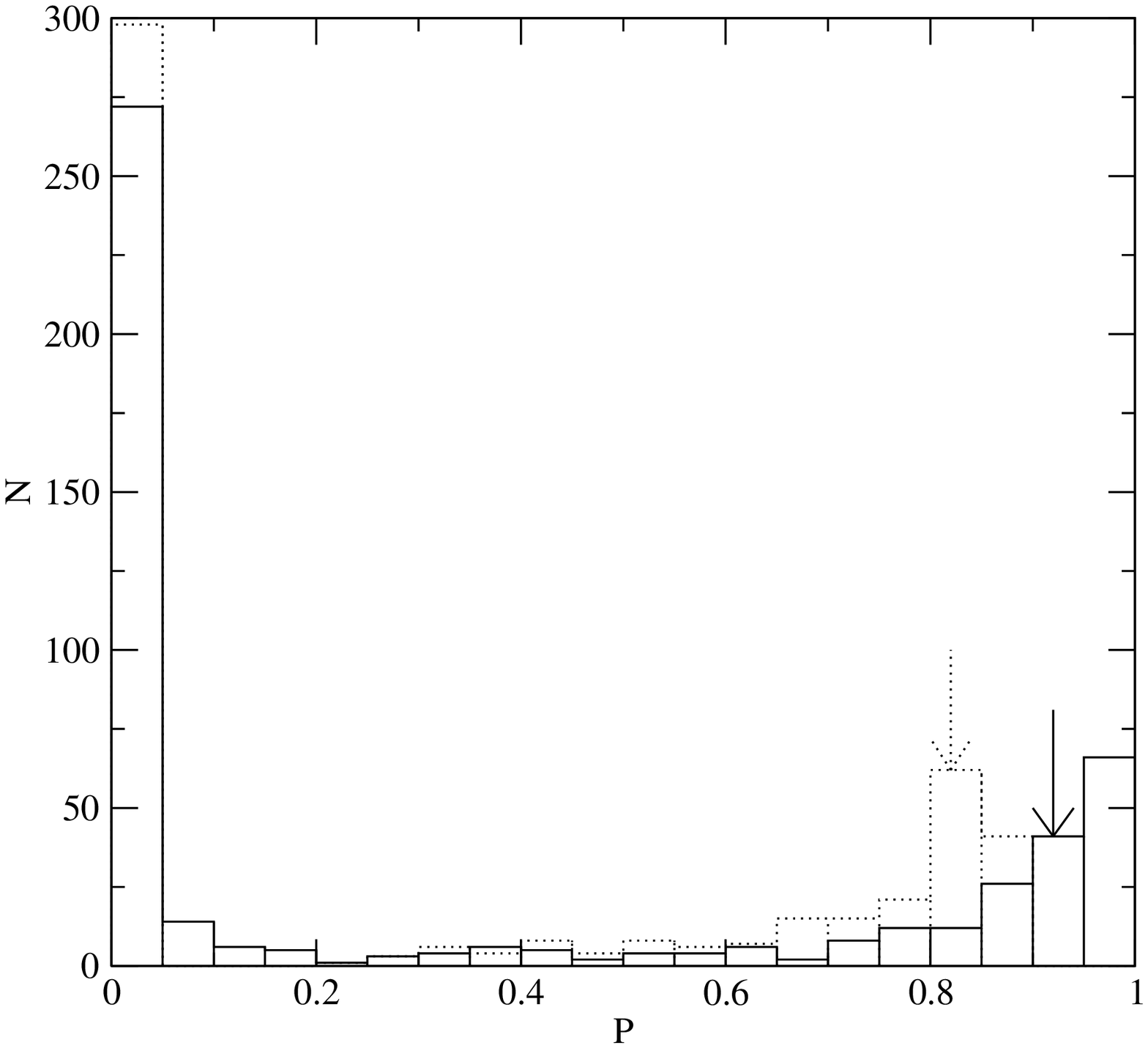}}
\end{center}
\caption{The histogram of cluster membership probability 
of NGC~2548. The solid line gives the results for traditional parametric 
method (Paper~I), while the dotted line corresponds to the non-parametric 
approach. The arrows mark the limiting probabilities for member
selection for each method.}
\label{fig7}
\end{figure}

	The cluster membership probability histogram (Fig.~\ref{fig7}) 
shows a clear separation between cluster members and field stars
in both approaches: the solid line is the traditional parametric method
(from Paper~I) while the dotted line is the non-parametric approach.
The non-parametric approach to cluster/field astrometric 
segregation gives us
an expected number of cluster members from the integrated volume of the 
cluster frequency function in the VPD areas of high cluster density, 
where PDF$\geq$ 3$\gamma$. 
This integration predicts that the sample contains 91 cluster members.
Sorting the sample in order of decreasing 
non-parametric membership probability, $P_{NP}$, the first 91 stars 
are the most probable cluster members. The minimum value of the non-parametric 
probability (for the 91-st star) is $P_{NP}=82\%$. Table~5 lists
the $P_{NP}$ for the 501 stars. 

\addtocounter{table}{1}

There is not an equivalent rigorous way to decide where to set the 
limit among members and non-members in the list sorted in order of 
decreasing parametric membership probability, $P_P$. But, if we accept 
the size of the cluster predicted by the non-parametric method, 91 
stars, we can consider that the 91 stars of highest $P_P$ are the most 
probable members, according to the results of the parametric 
technique. The minimum value of the parametric probability from Paper~I
(for the 91-st star) is $P_P=92\%$.

With these limiting probabilities ($P_{NP}\geq0.82$; $P_P\geq0.92$), we get 
a 91$\%$ (458 stars) agreement  
in the segregation yield by the two methods. 
The 43 remaining stars (9$\%$) 
with contradicting segregation should be carefully studied.
Discrepancies among the two approaches are 
actually expected due to the statistical nature of the methods themselves. 

As in Balaguer-N\'u\~nez et al.\ (\cite{Bal04a}), to set up a final and 
unique list, and trying not to reject true members, we accept as 
probable members of this cluster 
those stars classified as members by at least one of the two methods. 
This is equivalent to merging both lists (each with 91 stars) and 
eliminating duplicated entries. This way we get a list of 118 probable 
member stars.

As commented on the Introduction, the available information on 
radial velocities is not accurate nor complete enough to be useful in 
improving the membership segregation.

We have studied twenty stars in the 
area of the secondary clump on the tidal tail proposed by 
Bergond et al.\ (\cite{Bergond}).
Only three of them appear to be cluster members. The 
rest are randomly distributed in the VPD. 
Because our limiting magnitude for astrometry $B_{\rm inst}$$\approx$14
is brighter than the $B_{\rm cut}$$\approx$14.8 of 
Bergond et al.\ (\cite{Bergond}), we cannot discard the reality
of this clump. Unfortunately, our photometry does not cover this area. 
Deeper studies will be necessary.

\begin{figure*}
\resizebox{17cm}{!}{\includegraphics{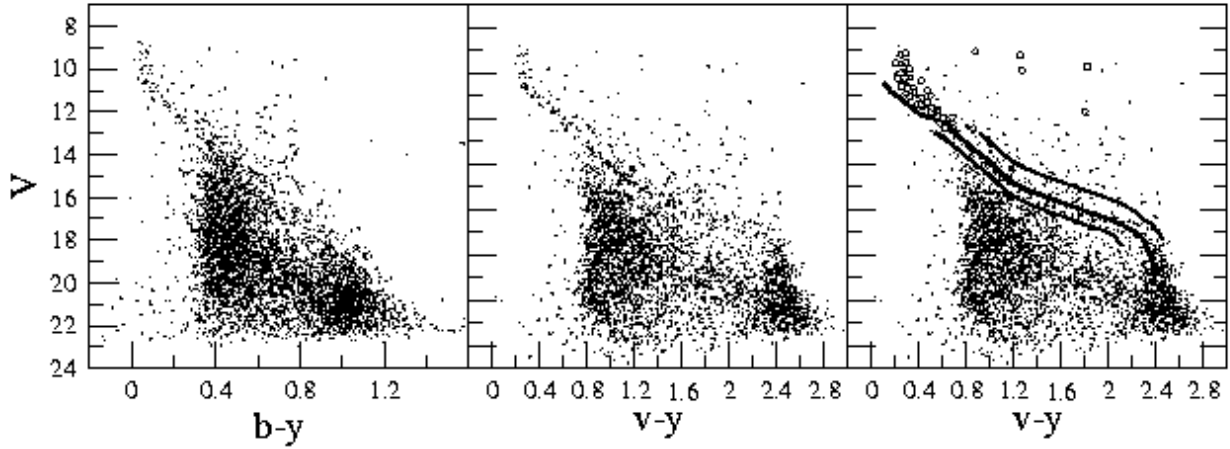}}
\caption{The colour-magnitude diagram of the NGC~2548 area.
Empty circles in the right figure are the astrometric members. 
Thick line is a shifted ZAMS, with the chosen margin for candidate members
($V+0.5$,$V-1$) in thin lines. See text for details.}
\label{HR}
\end{figure*}


\section{Colour-Magnitude diagrams}

	We use the $V vs\ (v-y)$ colour-magnitude diagram 
for our study. The colour-magnitude diagram based on this
colour index defines the main sequence of a cluster significantly better 
than the traditional $V vs\ (b-y)$ diagram (Fig.~\ref{HR} left and centre). 
The colour-magnitude diagram of all the stars in the area displays a
fairly well-defined main sequence.

\begin{figure}
\resizebox{\hsize}{!}{\includegraphics{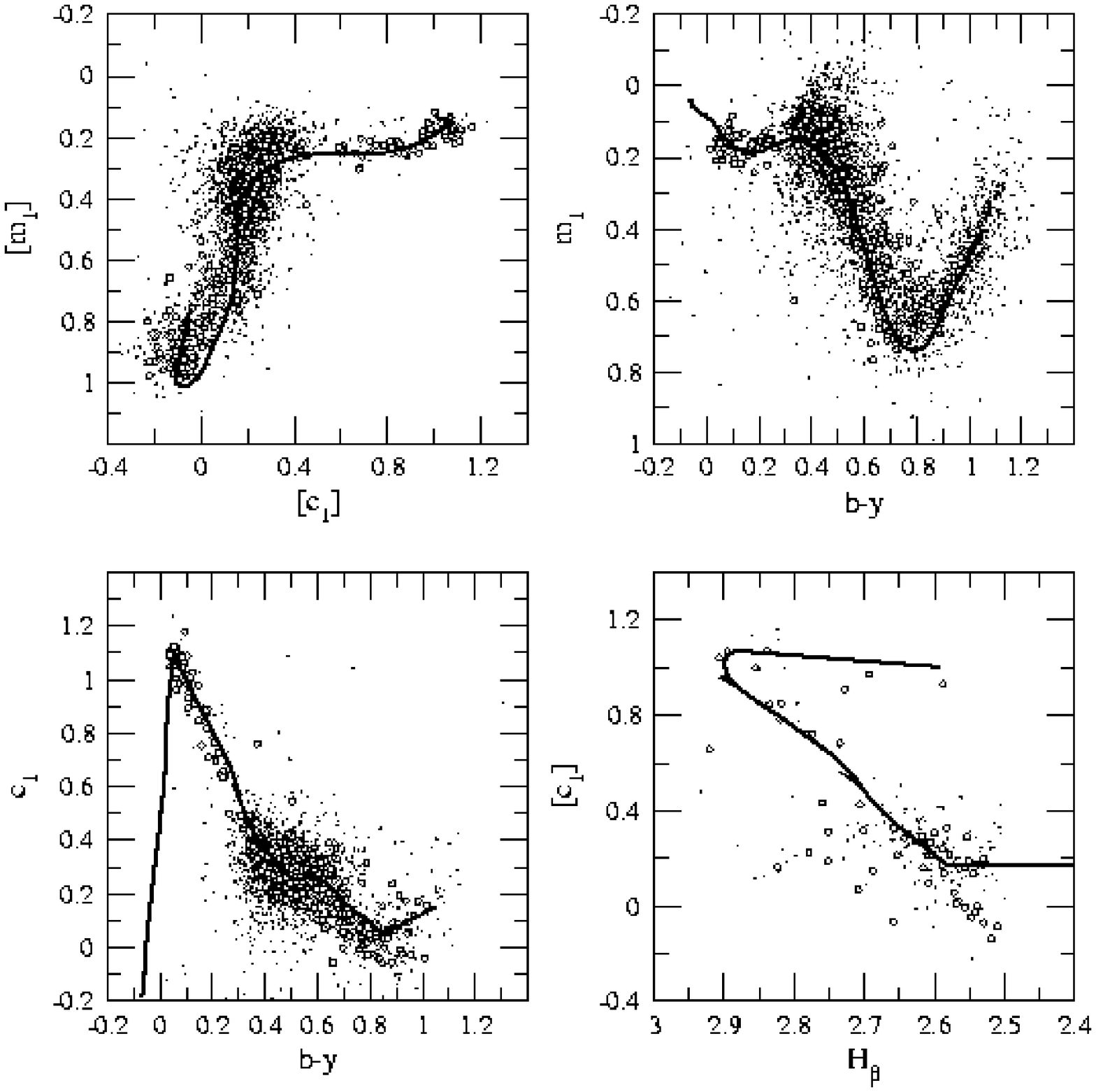}}
\caption{The colour-colour diagrams of NGC~2548.
           Empty circles denote candidate members of NGC~2548,
chosen with astrometric and non-astrometric criteria as explained
in Sect.~4.1. 
           The thick line is the standard relation shifted by 
 $E(b-y)$ = 0.06 when necessary.} 
\label{colour}
\end{figure}

\subsection{Selection of candidate member stars}

Our photometric 
measurements help to reduce the possible field contamination in the
proper motion membership among bright stars,
as well as to enlarge the selection of members towards faint magnitudes. 
Among those astrometric member stars with photometric measurements,
we find one star that is not compatible with the sequence of the
cluster outlined in the colour-magnitude diagram.
Our astrometric segregation of member stars has a limiting magnitude 
of $V\approx$~13. 
From this point down to $V=18$ we construct a ridge line
following a fitting of the   
observational ZAMS (Crawford \cite{Craw75}, \cite{Craw78}, \cite{Craw79}, 
Hilditch et al.\ \cite{Hil83}, Olsen \cite{Ols84})  
in the $V - (v-y)$ diagram. 
A selection of stars based on the distance to this ridge line is then
obtained. The chosen margin for candidates includes all the stars 
between $V+0.5$ and $V-1$ from the ridge line, as shown in the
right panel of Fig.~\ref{HR}. The margins were chosen to account for
observational errors and the presence of multiple stars.

This preliminary photometric selection is refined in the colour-colour 
diagrams (Fig.~\ref{colour}) with the help of the standard relations from the
same authors.  
A final selection of 331 stars in the area is plotted in Fig.~\ref{colour}
as empty circles in the $[m_1] - [c_1]$, $m_1 - (b-y)$, $c_1 - (b-y)$ and 
$[c_1] - H_{\beta}$ diagrams. 


\section{Physical parameters of the cluster}

The stars selected as possible cluster members were
classified into photometric regions and their physical parameters 
were determined as explained in Balaguer-N\'u\~nez et al. (\cite{Bal04b}). 
The algorithm, described in Masana et al.\ (\cite{Mas}) and Jordi 
et al.\ (\cite{Jordi97}), uses 
$uvby-H_{\beta}$ photometry and standard relations among colour indices
for each of the photometric regions of the HR diagram. 
Absolute magnitude, effective temperature and gravity as well as the
corresponding reddening, distance modulus, metallicity and raw 
spectral type and luminosity class are calculated for each star.
Typical errors are 0.25~mag in $M_V$, 0.15~dex in [Fe/H],
270~K in $T_{\rm eff}$, 0.18~dex in $\log g$ and 0.015~mag in $E(b-y)$.  

\begin{figure}
\resizebox{\hsize}{!}{\includegraphics{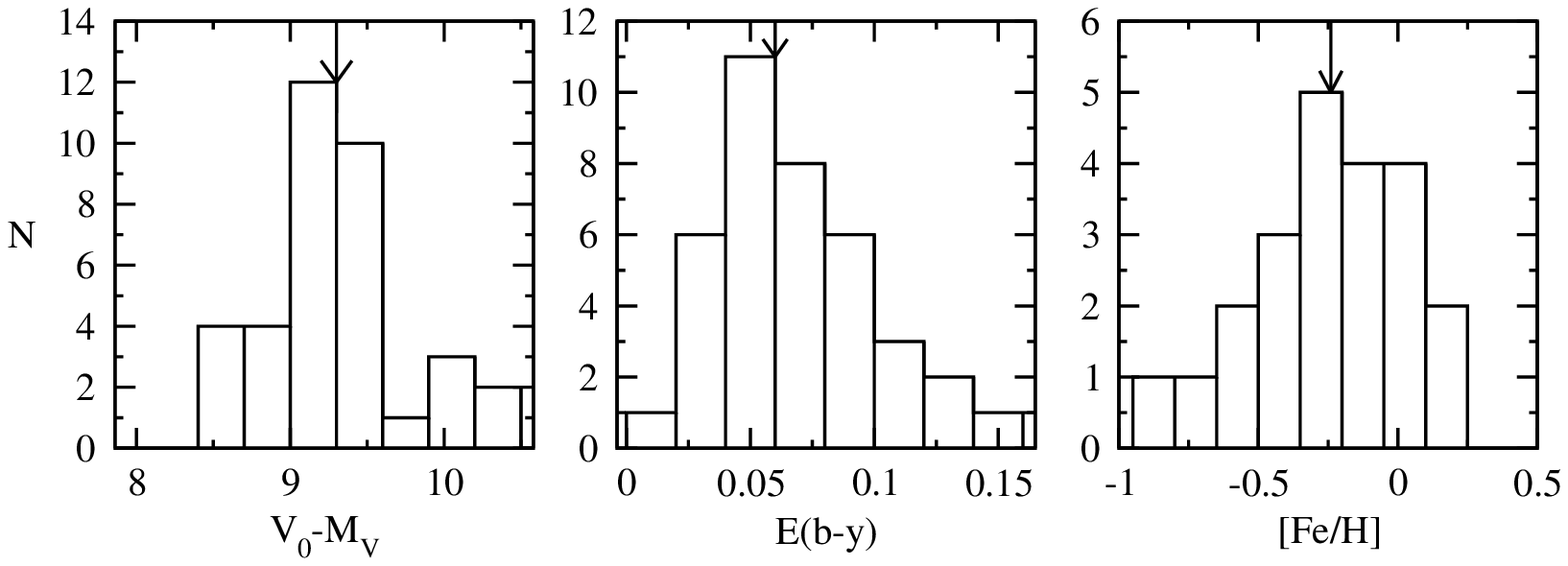}}
\caption{The histograms of the distance modulus, reddening and 
metallicity of the selected member stars of NGC~2548 with $H_{\beta}$
measurements. The arrows indicate the mean values adopted for 
the cluster.} 
\label{pf}
\end{figure}

Only 62 stars among the 331 candidate members have $H_{\beta}$
measurements. Thus, the computation of physical parameters is only 
possible for that subset.   
        The results are shown in Fig.~\ref{pf}. 
Excluding peculiar stars and those with inconsistencies in their 
photometric indices and applying an average with a  
2$\sigma$ clipping to that subset, 39 stars remain.
We found a
reddening value of $E(b-y)$~=~0.06$\pm$0.03
(corresponding to $E(B-V)$~=~0.08)
and a distance modulus of $V_0-M_V$~=~9.3$\pm$0.5 (725~pc, i.e.
about 200~pc above the galactic plane). 
Metallicity is better calculated studying only the 26 F and G stars in 
our sample following Masana et al.\ (\cite{Mas}). 
We found a value of [Fe/H]~=~$-$0.24$\pm$0.27. 

\begin{table*}
\begin{center}
\caption {Red giants in NGC~2548, with their membership. 
The first five are from the WEBDA list of Red Giants. 
Three of them are known spectroscopic binaries.
         }
\begin {tabular} {ccccccccccc}
\hline
 Phot. Id.  & WEBDA & $(b-y)$ & $V$ &  $P_P$ & $P_{NP}$ & M/NM & \\
\hline
 
1006 & 0870 &  0.670$\pm$0.0185 &  9.712$\pm$0.0185 & 0.07 & 0.84 & M? &  \\
2556 & 1218 &  0.335$\pm$0.0370 &  9.862$\pm$0.0305 & 0.83 & 0.85 & M &  \\
2732 & 1260 &  0.370$\pm$0.0002 &  9.050$\pm$0.0047 & 0.97 & 0.84 & M & SB \\
2894 & 1296 &  0.504$\pm$0.0004 &  9.215$\pm$0.0023 & 0.94 & 0.86 & M & SB \\
 (1) & 1560 &  0.724$\pm$0.0020 &  8.199$\pm$0.0060 & 0.88 & 0.84 & M & SB \\
3971 & 1521 &  0.312$\pm$0.0079 &  9.671$\pm$0.0081 & 0.00 & 0.00 & NM &  \\
4557 & 1616 &  0.719$\pm$0.0045 &  9.890$\pm$0.0045 & 0.00 & 0.00 & NM &  \\
2645 & 1241 &  0.777$\pm$0.0072 &  9.565$\pm$0.0072 & 0.81 & 0.66 & NM? &  \\

\hline
\label{RedG}
\end{tabular}
\end{center}
 (1)~Photometry taken from Olsen (\cite{Ols})
\end{table*}

	We have examined the bibliography in search of bright stars 
in the area of NGC~2548 and have found only one 
compatible with being a member of the cluster: The
brightest star (WEBDA 1560), that was saturated in our 
data. We therefore took its values from Olsen (\cite{Ols}). 
The stars in the red giant region are detailed in Table~\ref{RedG}. 
Three of them are known spectroscopic binaries.
We are cautious about our values of star WEBDA 1218, as the known values 
of $V$ are 9.64 (Pesch \cite{pesch}) and 9.63 (Oja \cite{Oja}),
about 0.2~mag brighter.
Star WEBDA 0870 was considered to be an astrometric non-member by 
Ebbinghausen (\cite{Ebb}) and in 
Paper~I, while Clari\'a (\cite{Claria})
studied its weak CN peculiarity and opened the possibility of a cluster member 
with peculiar composition and proper motion. Star WEBDA 1241 was considered 
to be a cluster member in Paper~I. 

\begin{figure*}
\resizebox{17cm}{!}{\includegraphics{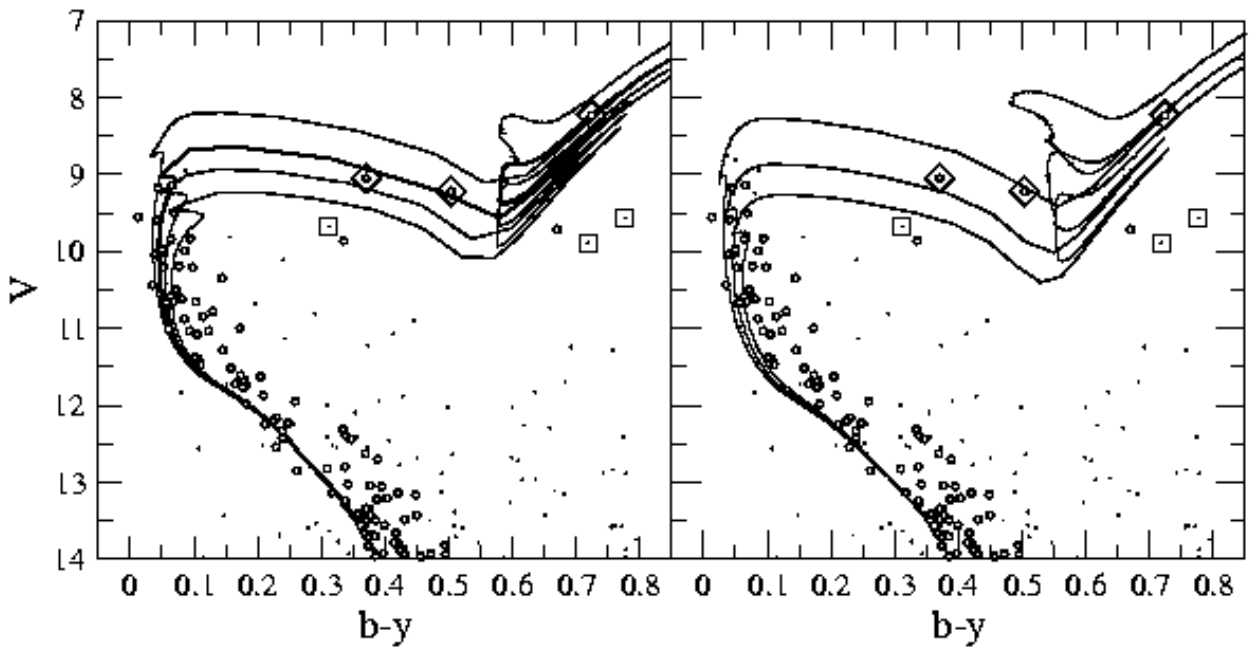}}
\caption{Isochrones from Pietrinferni et al.\ (\cite{Pie}) for
scaled solar models of [Fe/H] = $-$0.25 and ages of 300, 400,
500, 600 Myr for models with overshooting (left panel) and 200,
300, 400 Myr for canonical models (right panel). 
Empty circles are candidate members. 
Red giants are analyzed in detail (see text): diamonds are 
spectroscopic binaries, squares are astrometric non-members.
The adopted reddening and distance modulus are 
$E(b-y)$~=~0.06 and $V_0-M_V$~= 9.3.}
\label{iso}
\end{figure*}

	The recent publication by Clem et al.\ (\cite{Clem}) 
of empirically constrained colour-temperature relations in the Str\"omgren
system enables an isochrone fitting to our results. The best fit is
found for the Pietrinferni et al.\ (\cite{Pie}) tracks. Figure~\ref{iso} 
shows isochrones of [Fe/H]~=~$-$0.25 for canonical models (right panel) and
for models with overshooting (left panel). Taking into account the number of stars
before and after the MS hook, we found a best fit for models with
overshooting.
The estimated age is of 400$\pm$100 Myr ($\log t$~= 8.6). 
	Using a different set of tracks by
Schaller et al.\ (\cite{Scha92}) we also found agreement with a  
best estimation of the age of $\log t$ = 8.6$\pm$0.1.  
Neither Pietrinferni et al.\ (\cite{Pie}) nor Schaller et al.\ (\cite{Scha92})
isochrones provide a perfect fit to the giant members. Similar discrepancies were
found by Lapasset et al.\ (\cite{Lapasset}) studying the 
intermediate-age cluster NGC~2539.

	Our results are consistent with previous studies. 
Pesch (\cite{pesch}) gives an $E(B-V)$~= 0.04$\pm$0.05 and a distance of 630~pc
from $UBV$ photoelectric observations of 30 stars. Clari\'a (\cite{Claria})
studied $DDO$ photometry of 5 giant stars and gives a value 
of $E(B-V)$~= 0.06 and a distance of 530~pc.
Twarog et al.\ (\cite{Twa}) revised those values and gave an $E(B-V)$~=~0.05,
a distance of 590~pc and [Fe/H]~=~0.08 from 3 stars. 
Harris (\cite{Harr76}) gives a value of $\log t$~= 8.28 
from an analysis of nine stars. H\a{o}g \& Flynn (\cite{Hog}) give  
[Fe/H]~= $-$0.13, $E(B-V)$~=~0.04 and $\log t$~=~8.50 from $DDO$
photometry of three giants. 
Rider et al.\ (\cite{Rider}) fix a value $E(B-V)$~= 0.03 from bibliography 
and give values of distance
700~pc, $[Z/Z_{\sun}]=$~0.0 and an age of 400~Myr by isochrone fitting to 
$u'g'r'i'z'$ photometry. The recent study by Wu et al.\ (\cite{Wu05})
gives a slightly larger distance (780~pc) and a lower age (0.32~Gyr) also 
fitting isochrones with $E(B-V)$~= 0.04 and solar metallicity. 
A summary of this information 
is given in Table~\ref{resultcomp} for clarity.

\section{Gaps in the Main Sequence}

	Gaps in the main sequence have been observed and debated for 
a long time. 
In the Hyades, de Bruijne et al.\ (\cite{Bru})
confirmed the reality of two gaps near $B-V \sim$~0.38 and $B-V \sim$~0.48 
($T_{\rm eff}$ $\sim$ 6400~K). Evidence for the existence of gaps among field 
stars has also been debated (Newberg \& Yanny \cite{New}), and recently
Kovtyukh et al.\ (\cite{Kov}) showed a new gap among field stars
in the effective temperature range 5560 to 5610~K.

\begin{table*}
\begin{center}
\caption {Comparison of the physical parameters of the cluster with
results from other authors. See the text for detailed information of
the assumptions made in each case.}
\begin {tabular} {cccccc}
\hline
 Reference &$E(B-V)$ &  [Fe/H] &$\log t$ & Distance & Comments \\
    & mag & dex &  & pc & \\
\hline
This work & 0.08 &$-$0.24$\pm$0.27 & 8.6$\pm$0.1 & 725 & \\
Wu et al.\ (\cite{Wu05}) & 0.04 & 0.0 & 8.51 & 780 & Isochrone fit\\ 
Rider et al.\ (\cite{Rider}) &  & 0.0 & 8.60 & 700 & Isochrone fit \\ 
H\a{o}g \& Flynn (\cite{Hog}) & 0.04 & $-$0.13 & 8.50 &  &$DDO$
Phot. 3 stars\\ 
Harris (\cite{Harr76})& &  & 8.28 &  & Phot. 9 stars\\ 
Twarog et al.\ (\cite{Twa}) & 0.05 & 0.08 &  & 590 & $DDO$
Phot. 3 stars \\ 
Clari\'a (\cite{Claria}) & 0.06 & 0.1 &   & 530 & $DDO$
Phot. 5 stars\\ 
Pesch (\cite{pesch}) & 0.04 & &  & 630  & $UBV$ Photoelectric 30 stars\\ 
\hline
\label{resultcomp}
\end {tabular}
\end {center}
\end {table*}

\subsection{Method}

The characteristics of our photometry and member selection procedure
allow us to pursue the search for gaps in the main sequence not only
for the previously discussed cluster, NGC~2548 (M~48), but also in the
other clusters included in our project:
NGC~1817 (Balaguer-N\'u\~nez et al.\ \cite{Bal04a}, \cite{Bal04b}) and 
NGC~2682 (M~67, Balaguer-N\'u\~nez et al.\ \cite{Bal05}).  
To check the reliability of the results, we apply the same
gap search to the Pleiades (M~45) and the Hyades, two well-known
clusters in which gaps have been previously reported 
(de Bruijne et al.\ \cite{Bru}, 
Belikov et al.\ \cite{Bel}, 
and specially their Figs.~1 and 3, respectively.) 

	We estimate effective temperatures for our selection of 
candidate members following a new approach (Ribas et al.\ \cite{Ribas},
Masana et al.\ \cite{Mas05b}) 
based on fitting observed IR photometry with accurately calibrated 
synthetic photometry. We use our values of $V$ magnitude and the 
2MASS values for $J$, $H$, and $K$ magnitudes. 
The process requires metallicity and surface gravity data that we
compute from our Str\"omgren photometry as explained in previous sections.
The method is restricted to the temperature interval from 4000~K to 
8000~K, with quoted uncertainties of 0.5-1.3$\%$. 
The upper temperature limit is due to the increased dependence of the
results on the accuracy of $\log g$, while the lower limit is related 
to the decreasing performance of the models because of molecular bands. 
Ribas et al.\ (\cite{Ribas}) show that, in this range, the procedure
is essentially insensitive to the adopted value of [Fe/H] and $\log g$: 
uncertainties of 0.3~dex in metallicity and 0.5~dex in gravity 
induce $T_{\rm eff}$ deviations inferior to 0.5$\%$.
The number of member stars from our selection of clusters that also have
2MASS photometry in the range of temperatures under study is of 269 stars 
for NGC~2548, 307 stars for NGC~1817 and 588 stars for NGC~2682. For Pleiades 
we obtain 225 member stars and for Hyades 76.   

After getting the $T_{\rm eff}$ values, we proceeded to the
gap search following a method analogous to that proposed by
Rachford \& Canterna (\cite{Rach}). A simple $\chi^2$ test with 
one degree of freedom is used to evaluate the significance
of any candidate gap. To do this, we take a candidate gap
of width $W_{\rm in}$ and we compute the number of stars
within it, $N_{\rm in}$. This number is compared with the
stars located on both sides of the gap. We take two bins of 
equal width, $W_{\rm out}$, at the sides of the candidate gap, and
we count the stars inside them, $n_{\rm out}$. 
Then, $n_{\rm out}$ is
scaled to the size of the gap to give 
$N_{\rm out}=W_{\rm in}n_{\rm out}/(2W_{\rm out})$ and this quantity is
compared to $N_{\rm in}$ to get the $\chi^2$ value. 
To limit the effect of small number statistics,
and again following Rachford \& Canterna (2000), the computation is done
only when at least five stars are present on each side of the gap,
which guarantees that $n_{\rm out} \geq 10$.
	
We performed this calculation for a grid of gap widths 
75~K $< W_{\rm in}< 500$~K, at intervals of 25~K, and
placing the gap centres every 1~K over the intersection of
the temperature range covered by the method with the
range covered by the photometry. To prevent edge effects, intervals
of 300~K were avoided at the extremes of the intersection.

After several trials, we chose $W_{\rm out}=100$~K for our
photometry, but the Pleiades and Hyades required 
$W_{\rm out}=200$~K, due to the smaller size of the photometric
sample. It is important to note that widening $W_{\rm out}$
can alter the relative value of $\chi^2$ and its associated
probability, but with little effect on the significance of
the gap centre position or width.

Having computed $\chi^2$ for the whole range of gap centres and
sizes, we selected the local maxima of the resulting
table. We observed that changes in $W_{\rm in}$ do not 
make gaps appear or disappear, but only affect their significance.
All significant gaps have probability values higher than 0.99.

\begin{table*}
\begin{center}
\caption {Gaps in temperature empirically detected in the main sequence of open clusters.}
\begin {tabular} {cccccccccccccc}
\hline
   Cluster  & Age & [Fe/H] & Centre& Width& Centre& Width& Centre& Width& Centre& Width \\
   & Gyr & dex & K & K & K & K & K & K & K & K\\
\hline
Pleiades (M~45) & 0.1 & -0.03 & 7377& 300 & 6821  & 325 & 5577& 275 & 4845 & 275\\ 
NGC~2548 (M~48) & 0.4 & -0.24 & 7099& 475 & 6305  & 425 & 5465  & 350 & 4785 & 350\\
Hyades   & 0.7 &  0.15 & 7006& 475 & 6427& 250 & 5452  & 225 & 4972   & 275\\
NGC~1817  & 1.1 & -0.34 & 7323& 175 & 6674  & 350 & 5701  & 300 &    -   &   -\\
NGC~2682 (M~67) & 4.2 &  0.01 &    -  &   - &    -  &   - & 5593& 125 & 5017   & 200 \\
\hline
\label{gaps}
\end {tabular}
\end {center}
\end {table*}

\subsection{Results}

Table~\ref{gaps} gives the centres and widths ($W_{\rm in}$)
for the significant
gaps in the temperature range studied. The two well-known
B\"ohm-Vitense gaps are clearly detected in
Hyades and also Pleiades, which serves as a check of the
reliability of the method. Both gaps are significant
in the whole sample of clusters analysed, with the exception of M~67,
due to the much higher age of this stellar system (the main
sequence is so evolved that the corresponding effective temperature
range is no longer populated).

The gap recently reported by Kovtyukh et al.\ (\cite{Kov}) in field stars
at 5560-5610~K, and already suspected in the Hyades
(de Bruijne et al.\ \cite{Bru}), stands out not only in the
Hyades, but also in the other clusters surveyed.
In the case of M~67, Fan et al.\ (\cite{Fan}) also raised the
possible existence of this gap placed around 1~$M_{\odot}$. 

The reliability of the new gap detected around $T_{\rm eff} \sim 
4900$~K is stressed by its appereance in all the clusters under
study with the only exception of NGC~1817. The reason is
that, in this case, our
photometry does not reach stars colder than 5000~K.

Even though we know that some amount of field contamination is
present in our candidate member selection, it is worth noting that
the significance of the four gaps in our photometric data is outstanding.
If an even more reliable member selection would be possible, this could 
make the gaps even more evident in our data.

\begin{figure}
\resizebox{\hsize}{!}{\includegraphics{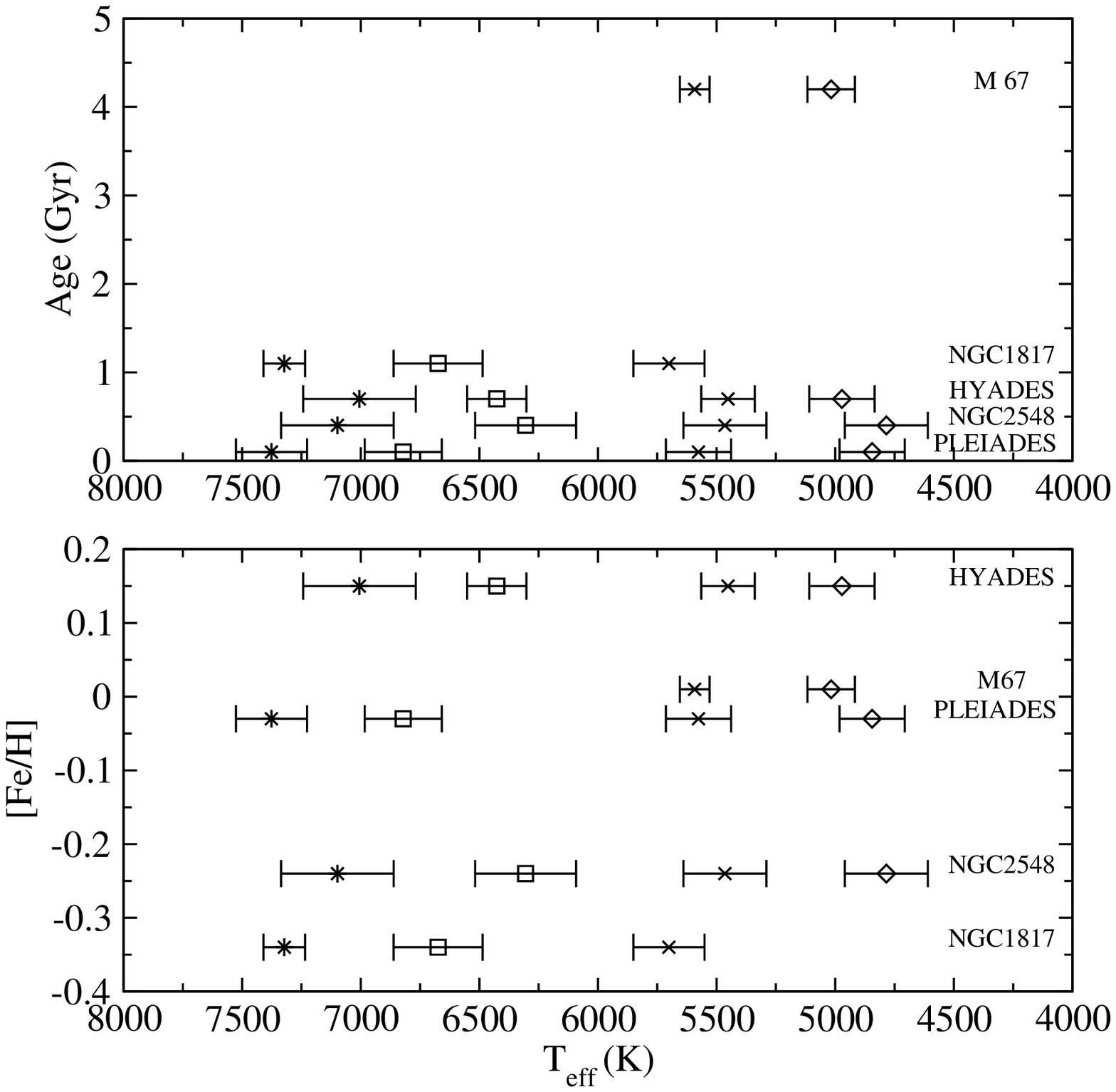}}
\caption{Gaps in temperature vs. age (upper panel) and metallicity 
(lower panel). See text for the metallicity uncertainties.
}
\label{Gapage}
\end{figure}

The position and width of the gaps are compared with
the age and metallitity of the clusters in Fig.~\ref{Gapage}.
No clear trend can be drawn from this comparison. However, metallicity 
effects could be masked due to the uncertainties in the metallicity 
determinations used for NGC~2548, NGC~1817 and NGC~2682 (M~67)
that are of the order of 0.2 dex.

Independently of the physical explanation for the existence of these 
gaps, there is no reason, in principle, to expect them to appear at 
the same positions and with similar sizes in all the 
clusters. There can be a complex dependency on metallicity, age and 
other parameters. However, the clusters studied show some regularity 
that allows to distinguish, as mentioned, four independent gaps. Even 
though we know that this kind of average may lack physical meaning, 
we give a table with mean, approximate parameters that caracterize 
the loci of the four gaps (Table~\ref{gaploci}). The colour 
indexes and masses are taken from standard relations (Schmidt-Kaler
\cite{schmidt-Kaler82}) and assume solar metallicity.

\begin{table}
\begin {center} 
\leavevmode
\caption{Orientative and approximate characterization of the mean loci 
of the gaps detected.}
{\scriptsize
\begin {tabular} {ccccc}
\hline
$<T_{\rm eff}>$~(K)  &  $\sigma_{<T_{\rm eff}>}$~(K) & {\it B--V} &
Spec. & Mass ($M_{\odot}$) \\
\hline
7200 & 180 & $\sim$0.30 & F0    &  $\sim$1.6 \\
6600 & 230 & $\sim$0.40 & F2-F5 &  $\sim$1.3 \\
5600 & 100 & $\sim$0.70 & G5-G8 &  $\sim$0.80 \\
4900 & 110 & $\sim$0.90 & K2    &  $\sim$0.75\\
\hline
\end {tabular}
}
\label{gaploci}
\end {center}
\end{table}

	Several theoretical explanations have been proposed for the 
three hotter gaps (see the references already given). Most
but not all, point to the need to improve the treatment of rotation
and convection in stellar models. However, theoretical and/or
empirical references to the fourth, colder gap are lacking. Spectroscopy
could confirm or reject the reality of this gap whose
significance is similar to that of the others, on the basis of our
photometric data.

\section{Conclusions}

In this paper we give a catalogue of accurate Str\"omgren
$uvby-H_{\beta}$ and J2000 coordinates for 4806 stars in an area of
34$^{\prime}$$\times$34$^{\prime}$ around NGC~2548. 

	We give a selection of probable members of NGC~2548, combining the
photometric study with an astrometric analysis using parametric as well
as non-parametric approaches. 
An improved determination of the physical parameters of this cluster 
based on our photometry gives: 
$E(b-y)$~= 0.06$\pm$0.03, [Fe/H]~= $-$0.24$\pm$0.27, 
a distance modulus of $V_0-M_V$~= 9.3$\pm$0.5 and an age of
$\log t$~= 8.6$\pm$0.1. 
The values are consistent with
previous studies (see Chen et al.\ \cite{Chen} and references therein). 

We perform a search for possible gaps in the main
sequence of this cluster and, also, of the other clusters
analysed in our project (NGC~1817, M~67). The Pleiades and
Hyades are added as a check and to construct a sequence of
ages and metallicities. We find four gaps in the
effective temperature interval 4000--8000~K. The coldest
gap, around 4900~K, had not been previously reported.

\vspace{3mm}
%

\

\begin{acknowledgements}

    We are grateful to Eduard Masana for very useful discussions and for
    providing the code for calculating effective temperatures.     
    We would like to thank Simon Hodgkin and Mike Irwin for their 
    inestimable help in the reduction of the images taken at the WFC-INT. 
    L.B-N. also wants to thank Gerry Gilmore and Floor van Leeuwen 
    for their continuous help and 
    valuable comments, as well as all the people at the IoA (Cambridge)
   for a very pleasant stay. L.B-N. gratefully acknowledges financial 
    support from EARA Marie Curie Training Site (EASTARGAL) during her 
    stay at IoA.
Based on observations made with the INT and JKT telescopes operated on 
the island of La Palma by the RGO in the Spanish Observatorio del Roque 
de Los Muchachos of the Instituto de Astrof\'{\i}sica de Canarias, and  
with the 1.52~m telescope of the Observatorio Astron\'omico Nacional (OAN) 
and the 1.23~m telescope at the German-Spanish Astronomical Center,
Calar Alto, operated jointly by Max-Planck Institut f\"ur Astronomie and 
Instituto de Astrof\'{\i}sica de Andalucia (CSIC).
    This study was also partially
    supported by the contract No. AYA2003-07736 with MCYT.
    This research has made use of Aladin, developed by CDS,
    Strasbourg, France.
This publication makes use of data products from the Two Micron All Sky Survey, 
which is a joint project of the University of Massachusetts and the Infrared 
Processing and Analysis Center/California Institute of Technology, funded by 
the National Aeronautics and Space Administration and the National Science
 Foundation.
\end{acknowledgements}

\end{document}